\documentclass[preprint,number]{elsarticle}

\usepackage{graphicx} 
\usepackage{amssymb}
\usepackage{lipsum}
\usepackage{bm}
\usepackage{subcaption}
\makeatletter
\def\ps@pprintTitle{%
   \let\@oddhead\@empty
   \let\@evenhead\@empty
   \let\@oddfoot\@empty
   \let\@evenfoot\@empty
}
\makeatother

\begin{document}

\begin{frontmatter}

\title{On Entropic Gravity from BFSS Matrix Theory}


\author{Korin Aldam-Tajima\fnref{footnote_label2} and Vatche Sahakian\fnref{footnote_label2}}
\fntext[footnote_label2]{Harvey Mudd College, Physics Department,  Claremont, CA 91711, \texttt{kaldamtajima@g.hmc.edu}, \texttt{sahakian@hmc.edu}}

\begin{abstract}

We study Matrix theory at strong coupling in a setting describing two static objects a fixed distance apart, using numerical techniques. We reproduce the exact general relativistic force law between the two objects as an entropic force in Matrix theory. This involves employing an operator algebra that represents an external observer measuring the relative positions and momenta of the two objects. We identify the location of the horizons of the objects from this entropic force and are led to a modification of the Schwarzschild spacetime inside the horizon. We find evidence that the inside of a black hole is instead to be described by AdS space. The conclusions constitute numerical validation of Verlinde's entropic gravity proposition and the fuzzball paradigm.
\end{abstract}

%

\end{frontmatter}

\section{Introduction and Highlights}
\label{sub:intro}

Matrix theory~\cite{Banks:1996vh,Berenstein:2002jq} is a natural computational playground for quantum gravity. It is posited to be a non-perturbative formulation of M-theory -- at strong coupling, its matrix degrees of freedom weave the fabric of M-theory space. 

Consider $SU(2)$ Banks-Fischler-Shenker-Susskind (BFSS) Matrix theory, describing Discrete Light-Cone Quantized (DLCQ) M-theory~\cite{Bigatti:1997gm,Susskind:1997cw,Seiberg:1997ad}, in a regime where the matrices $\hat{\bm{X}}^i$ (with $i=1,\ldots d-1$ being the target space index, and $d\leq 10$) can be arranged as follows
\begin{equation}
	\hat{\bm{X}}^i\sim \left(
	\begin{array}{ccc}
	\bm{X}_0^i & {\bm{X}}_R^i \\
	{\bm{X}}^{i\dagger}_R & -\bm{X}_0^i
	\end{array}
	\right)
\end{equation}
where $\bm{X}_0^i$'s describe a slowly evolving diagonal mode, while the off-diagonal $\bm{X}_R^i$'s describe much faster modes\footnote{We use bold lettering to denote quantum operators and hat for color matrix structure.}. We will refer to the longer timescale associated with the diagonal blocks as $\tau_D$, and the fast timescale as $\tau_R$. We are then studying the theory in a regime where
\begin{equation}
	\tau_D \gg \tau_R\ .
\end{equation}
This hierarchy of timescales is present in general when the separation between the  center of masses of the diagonal blocks -- in this SU(2) case simply the expectation value $X_0^i\equiv \langle\bm{X}_0^i\rangle$ -- is large, super-Planckian: we say we are in a regime that is associated with emergent space and gravity. The SU(2) instead of U(2) gauge group assures that we are in the center of mass frame of the system. This regime also corresponds to strong effective coupling in Matrix dynamics. In DLCQ M-theory language, we are focusing on two objects separated by a large distance $2\,|X_0|$, and the cloud of fast off-diagonal matrix modes encodes the effective interaction between the two objects. These objects can for example be two membranes wrapping some of the $10-d$ compact directions. 

To be more precise, we imagine an external observer is measuring the relative positions and momenta of the two M-theory objects represented by the diagonal mode. The relevant operator algebra that splits the system's Hilbert space through this measurement is given by\cite{Mazenc:2019ety,Balachandran:2013cq,DaltonGoold}
\begin{equation}
	\mathcal{A} = \left\langle\langle  | X \rangle_{D} \langle X'| \otimes \bm{1}_R\right\rangle\rangle
\end{equation}
where the subscript $D$ signifies the sector of the Hilbert space acted upon by operators in the direction of the third  color in the Pauli basis
\begin{equation}\label{eq:tau3}
	\hat{\tau}^3 = \left(\begin{array}{cc}1 & 0\\0 & -1\end{array}\right)\ \ \ ,\ \ \ \hat{\bm{X}}^i\rightarrow\left(\begin{array}{cc}\bm{{X}}_0^i & 0\\0 & -\bm{X}_0^i\end{array}\right)\ ;
\end{equation}
and the $R$ subscript indicates the subspace of the off-diagonal modes in the directions of $\hat{\tau}^1$ and $\hat{\tau}^2$ colors\footnote{This measurement process naturally breaks the gauge symmetry. In~\cite{sahakian2025} and the main text, we describe how this is to be handled.}. We propose that the measurement collapses the quantum state in the matrix diagonal sector into a coherent state $|\alpha \rangle_D$ where
\begin{equation}
	\alpha^i=\frac{1}{\sqrt{2}}\left(\xi^{-1} {X}_0^i+i \xi {P}_0^i\right)
\end{equation}
with the constant $\xi$ defined as\footnote{A coherent state is one with $\Delta X \Delta P=\mbox{ minimum }$ and maximal entropy while the first and second moments of position and momentum are fixed, described by a gaussian wavefunction. It is known to represent well the result of a typical experimental measurement with $\Delta X\Delta P\sim 1$~\cite{nickwheeler}.}
\begin{equation}
	\xi\equiv\sqrt{\Delta X / \Delta P}\ ,
\end{equation}
the ratio of uncertainties in position and momentum measurements. The coherent state  makes sure that we are measuring realistically two objects of finite width in space and momentum. Each object carries one unit of light-cone momentum. We will take $\Delta P\simeq \Delta X^{-1}$ and fix $\Delta X$ to be small enough in relation to $X_0$ to consider a reasonably precise measurement.

Importantly, we posit that the density matrix resulting from the measurement of the slow diagonal mode takes the form~\cite{sahakian2025}
\begin{equation}\label{eq:dm}
	\bm{\rho} = | \alpha \rangle_D \langle \alpha | \otimes \bm{R}_\alpha
\end{equation}
where
\begin{equation}
	\bm{R}_\alpha = \frac{\bm{Z}_\alpha^T}{\mbox{Tr}_R \bm{Z}_\alpha^T}
\end{equation}
with\footnote{This density matrix is normalized $\mbox{Tr}\, \bm{\rho} = 1$ and mixed $\mbox{Tr}\, \bm{\rho}^2 < 1$. And if we were to trace over the fast modes, we get $\bm{\rho}_D = \mbox{Tr}_R\, \bm{\rho} = | \alpha \rangle \langle  \alpha |$ which is a mixed state due to the over-completeness of the coherent states. }
\begin{equation}\label{eq:Z}
		\bm{Z}_\alpha^T = \exp{\left\{-\frac{1}{T} \bm{H}_R^\alpha \right\}}\ .
\end{equation}
Here, we define
\begin{equation}\label{eq:Heff}
	\bm{H}_R^\alpha \equiv \mbox{}_D\langle \alpha|\bm{H}_R|\alpha\rangle_D\ ,
\end{equation}
the expectation value of the part of the Hamiltonian of the system -- involving the fast modes and their couplings to the slow modes, -- in the coherent state of the diagonal mode. This is the effective Hamiltonian for the fast off-diagonal modes in the background of the slow diagonal one in an adiabatic regime. 
We are then saying that the fast modes in the $\hat{\tau}^1$ and $\hat{\tau}^2$ color directions are thermalized by chaotic dynamics at some temperature $T$, with the slow $|\alpha\rangle_D$ state acting as a background to the dynamics of the fast modes. $T$ here should be viewed as a parameter set by the two objects and tunes the amount of entanglement between the slow mode and the fast modes after the measurement\footnote{Note that the phase space of the diagonal mode is much larger than that of the off-diagonal ones since the objects can access infinite space while the off-diagonal modes will be confined in phase space.}. At finite temperature $T$, the supersymmetry in the BFSS Matrix theory is broken, but the $T\rightarrow 0$ limit recovers this symmetry. To capture general scaling relations and identify emergence of gravity, we will see that it is  sufficient to work in the {\em bosonic} Matrix theory at finite $T$ while dropping Fermionic terms; however, the $T\rightarrow 0$ limit will need input from supersymmetry, as is well-known from other works~\cite{Tafjord:1997bk}-\cite{Lee:2004kv}. 

Given this setup, let us first list the main results of this work before presenting some of the details. We numerically compute, at strong and weak coupling, the energy spectrum of~(\ref{eq:Heff}) and the entanglement entropy for the density matrix~(\ref{eq:dm}). We find:

\begin{itemize}
	\item At distances between the two objects of order the M-theory Planck length, the eigenvalue spectrum exhibits a marked transition. 
	\item At distances greater than the Schwarzschild radii of the two objects, we recover the gravitational force as an entropic force law. This includes the entire general relativistic gravitational force law for two static objects, not just the leading Newtonian piece, in Schwarzschild coordinates. 
	\item We identify the location of horizons of the two objects in Matrix theory consistently with general relativity. 
	\item Inside the horizon, we find diverging conclusions between the entropic force and general relativity, potentially signaling a breakdown of general relativity at super-Planckian distances, where horizons form.
	\item Inside the horizon, we see evidence of emergence of Anti-de Sitter (AdS) space, and a resolution of the singularity -- connecting the picture with holography in the AdS/CFT correspondence setting. 
	\item The complete picture that emerges is a validation of Verlinde's entropic gravity~\cite{Verlinde:2010hp} and the fuzzball proposal~\cite{Lunin_2001,Lunin_2002,Mathur_2005}. 
\end{itemize}

\vspace{0.3in}

{\bf Highlights}\\

These are a remarkable set of results. Let us dive into some of the details. The bosonic sector of the $SU(2)$ BFSS Matrix theory is described by the Hamiltonian
\begin{equation}\label{eq:BFFSH}
		\bm{H}=\frac{1}{2} {{\bm{{\dot{X}}}}^i_a}^2+\frac{g^2}{4} \bm{X}^i_a \bm{X}^j_b \bm{X}^i_a \bm{X}^j_b-\frac{g^2}{4} \bm{X}^i_a \bm{X}^j_b \bm{X}^i_b \bm{X}^j_a\ ,
\end{equation}
with $a,b=1,2,3$ being color indices. $\bm{H}_R$ introduced above is part of this Hamiltonian when expanded about the configuration given by~(\ref{eq:tau3}) (see main text for details). We will also set the dimension of the target space such that $i,j=1,\ldots d-1$ -- that is, we work in some framework of compactified M-theory. 

The coupling $g$ is dimensionful, related to M-theory parameters as
\begin{equation}
	g^2 = \frac{R_+^3}{4\pi^2\ell_P^6}
\end{equation}
where $R_+$ is the size of the light-cone direction and $\ell_P$ is the Planck length in five dimensions\footnote{In this notation, Newton's constant in five dimensional M-theory is $G_5=(2\pi)^2\ell_P^3$ and $\ell_P=(2\pi)^{4/3}(\ell_P^{(11)})^3/V_6^{1/3}$, where $V_6$ is the six dimensional volume of compactification. The M-theory-IIA duality map retains $R_+=g_s\ell_s$ and $\ell_P=g_s^{1/3}\ell_s$, where $R_+$ is the size of the light-cone direction, $g_s$ is the string coupling, and $\ell_s$ is the string scale.}.

Given a separation $2\,|X_0|$ between the two M-theory objects with zero relative momenta, we compute numerically the spectrum of the Hamiltonian~(\ref{eq:Heff}) and, fixing a temperature $T$, we study the thermodynamics of the cloud of off-diagonal modes in the background of the slow diagonal mode as a function of $T$. In~\cite{sahakian2025}, it was shown that the force experienced by the two M-theory objects is the entropic force given by
\begin{equation}
	{F}_{ent} = T\,{\nabla}S\ ,
\end{equation}
where $S$ is the entropy from~(\ref{eq:dm}), that of the off-diagonal entangled with the diagonal\footnote{\cite{sahakian2025} also demonstrates how the $T\rightarrow 0$ leads to a force arising from the gradient of the ground state energy of the system, consistent with previous results~\cite{Tafjord:1997bk}-\cite{Lee:2004kv}.}.
We numerically compute this force at various couplings in four spacetime dimensions. We also present a preliminary treatment of the three dimensional case\footnote{Recently, a perturbative approach was developed in a related context by~\cite{Weishui}.}.

Expanding the Hamiltonian~(\ref{eq:BFFSH}) about the configuration given by~(\ref{eq:tau3}), we identify the two timescales as  
\begin{equation}
	\tau_D \simeq X_0^2\ \ \ ,\ \ \ \tau_R \simeq \frac{1}{g\,X_0}\ .
\end{equation}
The latter appears as an oscillator frequency, confining the phase space of the fast modes. We then need  $\tau_D\gg\tau_R$, implying
\begin{equation}\label{eq:regime}
	g\,X_0^3\gg 1\ .
\end{equation}
In these conventions, we have~\cite{Martinec:1998ja,Sahakian:2000bg,polchinski}
\begin{equation}
	2\,|X_0| =  \frac{r}{\sqrt{R_+}}\ \ \ ,\ \ \ 
\end{equation}
where $r$ is now the physical separation between our two objects in M-theory units\footnote{Without loss of generality, we align our target space axes so that the two objects lie along one of the axes; so the radial direction is mapped onto one of the $X_0^i$ which we henceforth write as  $X_0$.}. We then have the statement $\tau_R\ll\tau_D$ transform into
\begin{equation}
	r \gg \ell_P
\end{equation}
So, the proposed hierarchy of timescales is saying that the two objects are separated by a distance much larger than the M-theory Planck length. This is indeed the regime we expect to identify with emergence of space. 

So, we have two scales in the problem: $g$ and $X_0$
\footnote{$\Delta X$ will be set to around 10\% of the minimum $X_0$ we use, and $\Delta P \simeq 1/\Delta X$.}. We can use $g$ as the unit of measurement and define the following dimensionless parameters:
\begin{equation}
	g_X\equiv  g^{1/3}X_0 = \frac{r}{2\,(2\pi)^{1/3}\ell_P}\sim \frac{r}{\ell_P}
\end{equation}
as a dimensionless coupling; and
\begin{equation}
	t \equiv T g^{-2/3}\ \ \ \epsilon = E\, g^{-2/3}
\end{equation}
as dimensionless temperature and energy respectively; and finally
\begin{equation}
	{f}_{ent} = {F}_{ent} g^{-4/3}
\end{equation}
for dimensionless entropic force.

\begin{figure}[h]
\centering
\includegraphics[width=0.75\columnwidth]{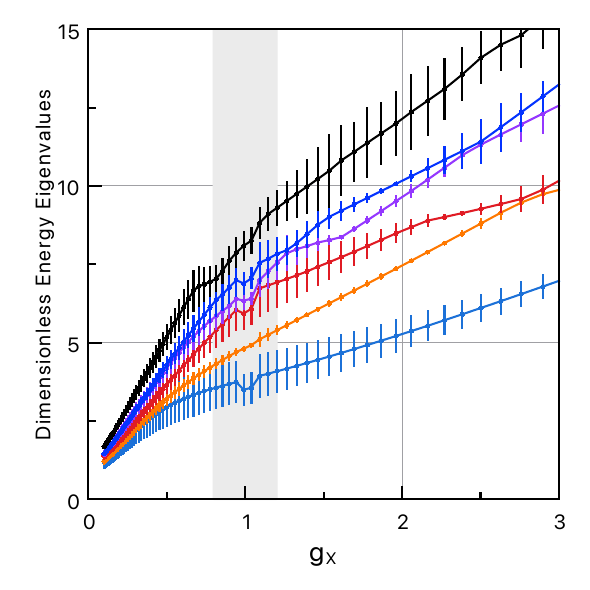}
\caption{\em Some of the low eigenvalues of the Hamiltonian~(\ref{eq:Heff}). The lowest line is the ground state. We also show conservative error bars estimated from the convergence of the numerics as a function of a UV cutoff. The shaded region, $g_X\sim 1$,  corresponds to the weak-strong coupling transition, and to the $r\sim \ell_P$ point.}\label{fig:eigenvalues}
\end{figure}
We start by numerically computing the eigenvalues and eigenstates of the Hamiltonian given by~(\ref{eq:Heff}) -- making sure we throw out states not killed by a residual Gauss constraint. 
Figure~\ref{fig:eigenvalues} shows the eigenvalue spectrum for $d=3+1$ as a function of $g_X$ with the use of a UV cutoff to manage the large Hilbert space. So, we see here some of the low-lying energy states only. We clearly witness a transition as we cross from weak to strong coupling, into $r>\ell_P$\footnote{Whether this is a phase transition or cross-over cannot be resolved in a finite system, far away from a thermodynamic limit.}. 

\begin{figure*}[h]
  \centering
  \includegraphics[width=0.9\columnwidth]{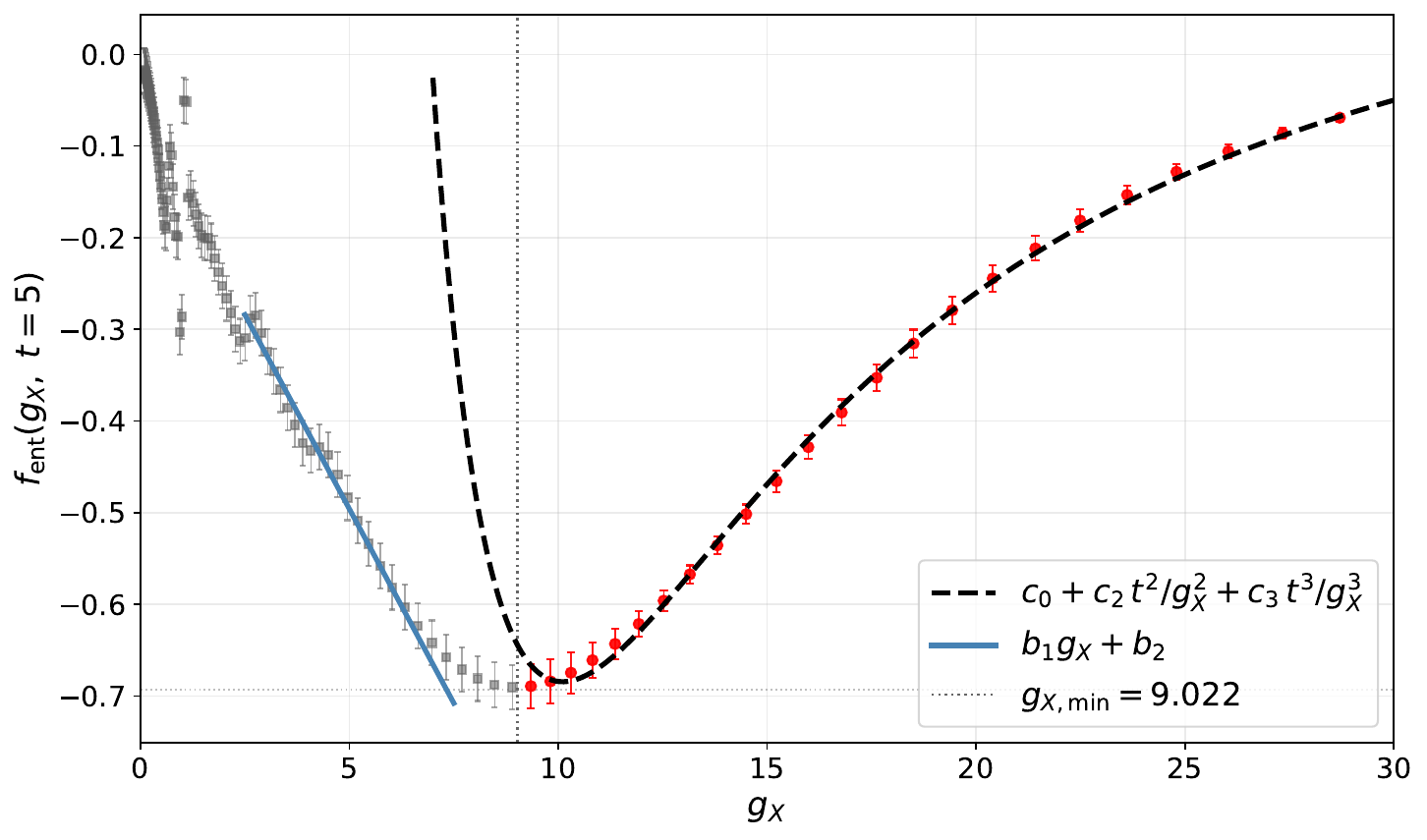}
  \caption{\em The entropic force as a function of $g_X\sim r/\ell_P$ for spacetime dimension $d=3+1$. At small but still super-Planckian distances, $f_{ent}\sim -r$; at larger distances, $f_{ent}$ is the gravitational force from general relativity, with a horizon at the minimum point of the force profile. In this plot, we show the case $t=5$, but the fit was generated across many values of $t$. The dashed curve is the result from general relativity fitted to the data.}
  \label{fig:entropicforce}
\end{figure*}
Next, in Figure~\ref{fig:entropicforce}, we show the entropic force experienced by the two objects computed from this energy spectrum. At strong coupling, we find
\begin{equation}\label{eq:fit}
  f_{\rm ent} = c_0 - c_2 \frac{t^2}{g_X^2} + {c_3}\frac{t^3}{g_X^3}\ ,
\end{equation}
with 
\begin{equation}
  c_0 = +0.178 \pm 0.002 \ \ \ ,\ \ \ 
  c_2 = +10.56 \pm 0.08 \ \ \ ,\ \ \ 
  c_3 = +14.2 \pm 0.2\ .
\end{equation}
This is a very robust fit, with 
\begin{equation}
  \chi^2_{\nu} = \frac{160.3}{217} = 0.739, \qquad
   N = 220\ .
\end{equation}
Contrast~(\ref{eq:fit}) with the exact gravitational force law from general relativity, for two momentarily static objects of mass $M$ each, that carry one unit of light-cone momentum each. At a distance $r$ apart and written in Schwarzschild coordinates, we have
\begin{equation}\label{eq:force}
	F_{grav} = M \frac{d^2r}{dt^2} = -\frac{G_4 M^2}{r^2}+\frac{2\,G_4^2M^3}{r^3}
\end{equation}
where $G_4 = G_5/(2\pi R_+) = 2\pi\ell_P^3/R_+$ is Newton's constant. To compare this to~(\ref{eq:fit}), we have to restore the right units using the M-theory-IIA duality map. We get from~(\ref{eq:fit})
\begin{equation}\label{eq:Fentscaled}
	F_{ent} = - 4\pi c_2\frac{\ell_P^3}{R_+}\frac{\mu^2 M^2}{r^2}+16\pi^2 c_3\frac{\ell_P^6}{R_+^2}\frac{\mu^3 M^3}{r^3}+\frac{R_+}{4\pi\,\ell_P^3}c_0\label{eq:gr}
\end{equation}
where we wrote $T=\mu\, M$, suggesting a mapping of the temperature parameter onto the mass of the objects. We also note that, in the relation between force and energy, there is one scale of {\em transverse} inverse length from the gradient of the energy, which scales as $\ell_P$ and not as $\ell_P^2/R_+$.

Now, it is already clear that the entropic force is capturing the scaling of the gravitational force with distance and mass, with the correct signs of the two terms. However, let us try to push things further and see how close the two expressions are numerically. To map~(\ref{eq:Fentscaled}) onto~(\ref{eq:force}), we need to consider several factors:
	
\begin{itemize}
	\item First, while the time coordinate in Matrix theory is to be mapped onto the time of the Schwarzschild metric -- identifying the observer measuring the two objects as located at asymptotic infinity, there is an ambiguity on how to map the $r$ coordinate arising in Matrix theory with respect to the radial coordinate of the Schwarzschild metric: in principle, there could be numerical scale factor between them. Looking at the form given by~(\ref{eq:fit}), we see such a scale factor can be absorbed in the parameter $\mu$, except that the force depends on it through acceleration -- an overall numerical factor that needs to be accounted for. We parameterize this ambiguity with a number we call $\gamma$ defined as
	\begin{equation}
		F_{ent}=\gamma^2 F_{grav}=\gamma^2\frac{d^2 r}{dt^2}\ \ \ ,\ \ \ r_{BFSS} = \gamma^2 r\ .
	\end{equation}
	
	\begin{figure}[h]
	  \centering
	  \includegraphics[width=0.65\columnwidth]{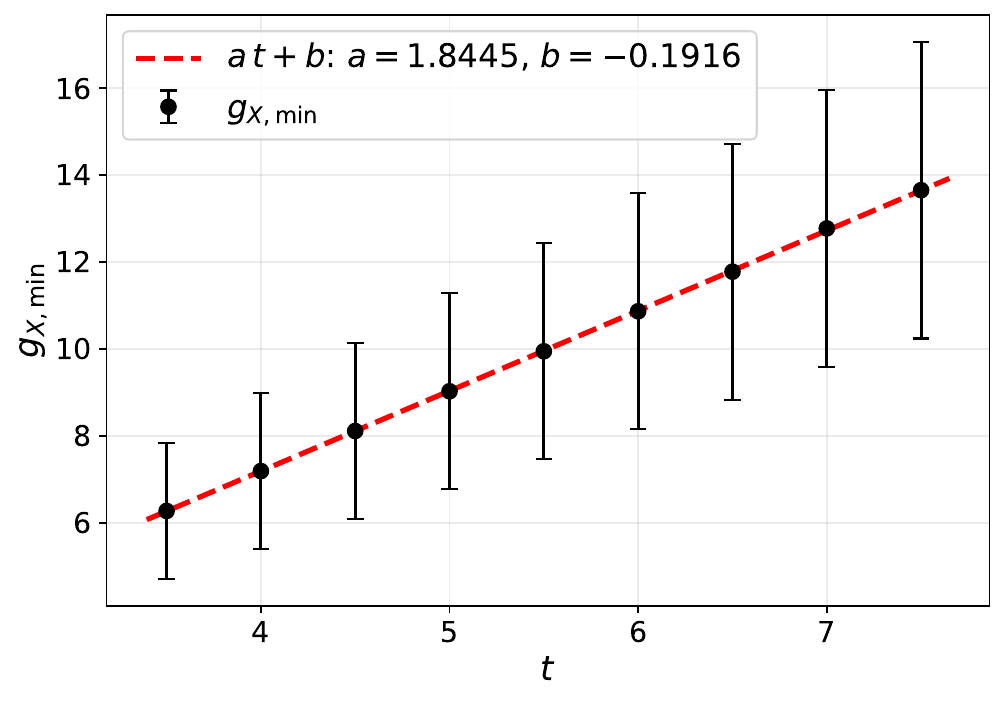}
	  \caption{\em The shift of $g_X$ at the minimum of the entropic force as a function of $t$. The large error bars are due to systematic errors from the matrix UV cutoff used.}
	  \label{fig:horizon}
	\end{figure}

	\item Next, by varying $t$ we identify the value of $g_{X,min}$ at the minimum of the entropic force that naturally maps onto the location of the horizon, as shown in Figure~\ref{fig:horizon}. We see a linear shift with temperature $t$ or equivalently mass as follows
\begin{equation}
	  g_{X,{\rm min}} = a\,t + b
	\end{equation}
	with
	\begin{equation}
	  a = 1.8 \pm 0.5\ \ \ ,\ \ \ 
	  b = -0.2 \pm 2.5\ .
	\end{equation}
	The error bars and corresponding errors on the fit parameters are large because of a matrix UV cutoff. Our numerics involve several sources of error and the full error analysis can be found in Appendix A. The bottom line is that, for the region outside the horizon, the statistical error estimates we have quoted for $c_0$, $c_1$, and $c_2$ earlier are robust. But very near the horizon, errors from the matrix UV cutoff we employ start to increase and dominate systematically at around 30\%. This is reflected in the large error bars in the figure, despite the solid linear form of the data arising from statistical self-consistency.   
	
	\item The ratio of the first term to the second in the force is
	\begin{equation}
		\mbox{ratio} = \frac{r}{2\,G_4 M}=\frac{r}{r_0} \rightarrow \frac{c_2 g_X}{c_3 t}  
	\end{equation}
	At the horizon $r=r_0=2\,G_4M$ or $g_X=g_{X,min}$, we then have
	\begin{equation}\label{eq:ratiotest}
		\mbox{ratio at $r_0$} = 1 = \frac{c_2 a}{c_3} = 1.4 \pm 0.4
	\end{equation}
	where we dropped the intercept $b$ which is consistent with zero.  

	\item The entropic force has a constant term $c_0$ not present in~(\ref{eq:force}). To reconcile this, we need to remember that we have dropped the fermions altogether from the supersymmetric BFSS theory in arriving at these results. These contributions should become more important as $t$ or the mass of the objects go to zero, and should become less important at high temperatures or large mass. The $c_0$ term makes this point very clear. The constant shift is an asymptotic force. A constant force corresponds to a linear term in $gX$ in the entropy and equivalently the free energy. In the partition function, large $g_X$ is equivalently $t\rightarrow 0$, where supersymmetry becomes important. This term is known to be {\em exactly} canceled by the leading fermionic contributions at $t=0$~\cite{Tafjord:1997bk}-\cite{Lee:2004kv}. So, the $c_0$ term is an artifact of having dropped the contributions of the fermions from the computation. More on this in the Conclusion section. All this suggests that exact numerical matching between the two expressions is not to be expected in general since the addition of the fermionic degrees of freedom is expected to contribute to the coefficients $c_0$, $c_2$, and $c_3$.
	
	\item Now, let us map~(\ref{eq:fit}) onto~(\ref{eq:force}) dropping the constant shift, and we find
\begin{equation}\label{eq:relations}
	\mu = \frac{\gamma}{\sqrt{2\,c_2}}\ \ \ ,\ \ \ \gamma^2=\frac{2\,c_2^3}{c_3^2}\ .
	\end{equation}

	\item Finally, we can now identify directly $g_{X,min}$ with the location of the horizon, taking into account of the $\gamma$ factor from the radial coordinate. We get
	\begin{equation}
		\gamma = \frac{\sqrt{2\,c_2}}{a}
	\end{equation}
\end{itemize}

Let us summarize these results. First, we emphasize that the Matrix entropic force captures the scaling and signs for the {\em exact} general relativistic force outside the horizon: we have checked that the coefficients of higher powers of $t/g_X$ are essentially statistically zero and other possible functional models do not fit well. Despite the fact that the fermions were dropped out of the computation, we can now also gauge how close is the match between the entropic force and gravity at the {\em numerical level}. We have 5 fit parameters; $c_0$, $c_2$, $c_3$, $a$ and $b$. $b$ is consistent with zero and we argued away $c_0$ due to the known role of the fermionic contributions. This leaves 3 fit parameters. But we have 2 arbitrary constants, $\gamma$ and $\mu$. So, the numerical comparison boils down to one test of numerical matching, which can be represented by equation~(\ref{eq:ratiotest}). This is a decent agreement bearing in mind that half of the degrees of freedom of BFSS Matrix theory were ignored and the temperature/mass parameter is only 7 to 8 times larger than the Planck scale.
   
Putting everything together, we conclude that {\em the Matrix entropic force captures the gravitational force from general relativity between two static objects in Schwarzschild coordinates.}

Note that this entropic force also predicts a breakdown of general relativity at the horizon, at distances much larger than the Planck length. Let us go back to Figure~\ref{fig:entropicforce} and consider small $g_X$, still with $r>\ell_P$, the region that we now realize corresponds to inside the horizon of the gravitating objects. In this region, the horizons of the two objects start intersecting. The numerical errors are more significant here too: another source of systematic error dominates the computation in this region, resulting from the truncation of the partition function. This is discussed in detail in Appendix A. So, the conclusions are somewhat less robust. We however can say that 
\begin{equation}
	f_{ent} = A(t) {g_X} \Rightarrow F_{ent}= \frac{R_+}{(32\pi^2)^{2/3}\ell_P^4}\,A(t)\,r\ ,
\end{equation}
with $A(t)<0$, consistent with a negative cosmological constant that depends on the mass: an attractive harmonic oscillator with the radial distance $r$. We do not have enough numerical accuracy to determine the scaling of $A(t)$ with $t$. $A(t)$ is related to the energy density or tension of an energy wall at the location of the horizon if we were to model the entropic force with general relativity: Schwarzschild outside the horizon and AdS inside, with a membrane at the horizon.  Irrespective, there is another remarkable conclusion here: the suggestion that general relativity breaks down at and inside the horizon, with the region inside best represented by AdS space. More on this in the Conclusion section.

The rest of the paper presents the details of these results, including aspects of the numerical techniques employed collected in the appendices. In the Conclusion section, we outline next steps and broader implications.

\section{Bosonic BFSS Quantum Mechanics}
\label{sub:bfss}

We consider the bosonic sector of the BFSS Matrix theory, described by the Hamiltonian~(\ref{eq:BFFSH})~\cite{Fuster:2005js}. We canonically quantize the system with
\begin{equation}
	\left[\bm{X}^i_a, \bm{P}^j_b\right]=i \delta_{a b} \delta^{i j}\ .
\end{equation}
and we impose the Gauss constraint on the Hilbert space arising from fixing the static gauge 
\begin{equation}
	\bm{G}_a=g\, \epsilon_{a b c} {\bm{X}^i}_b \cdot {\bm{P}^i}_c\Rightarrow \bm{G}_a |\Psi\rangle = 0\ ,
\end{equation}
The ground state of the system is defined by
\begin{equation}
	\bm{A}^i_a |\Psi_0\rangle=0
\end{equation}
where 
\begin{equation}
	\bm{A}_a^i=\frac{1}{\sqrt{2}}\left(\xi^{-1} \bm{X}_a^i+i \xi \bm{P}_a^i\right)\ .
\end{equation}
The role of the constant $\xi$ will become clear soon. A general state in the Hilbert space is constructed by
\begin{equation}
	|\Psi\rangle=M\left[\bm{A}^{\dagger}\right] |0\rangle .
\end{equation}
where
\begin{equation}
	M\left[\bm{A}^{\dagger}\right]=\sum_n \mu_{a_1 \ldots a_n} \bm{A}_{a_1}^{\dagger} \ldots \bm{A}_{a_n}^{\dagger}
\end{equation}
with the $\mu_{a_1 \ldots a_n}$'s being invariant SU(2) tensors.

The outcome of a measurement of the position and momentum of the slow degree of freedom is best represented by a coherent state~\cite{PhysRev.152.1103}
\begin{eqnarray}
\bm{D}_\alpha | 0 \rangle &=&\exp \left\{\alpha^i_a{\bm{A}^i_a}^{\dagger}-\bar{\alpha}^i_a\, \bm{A}_a^i\right\} | 0 \rangle \nonumber \\ &\equiv& | \alpha \rangle\ .
\end{eqnarray}
We assure that this state is physical by adding the following identification to the Hilbert space
\begin{equation}\label{eq:equivalence}
	|\alpha \rangle \approx | \beta \rangle\ \ \ \ \mbox{if ${\bar{\alpha}}_a^i \alpha_a^i = {{\bar{\beta}}}_a^i {\beta}_a^i$ (no sum over $i$, $\forall i$)}
\end{equation}
Without loss of generality, we can pick a representative of this equivalence class on the gauge orbit, say the one pointing in the $\tau^3$ direction. We write for notational purposes
\begin{equation}
	\bm{X}^i \equiv \bm{X}_3^i\ \ \ ,\ \ \ \bm{P}^i \equiv \bm{P}_{3}^i\ .
\end{equation}
We then have
\begin{equation}
	\bm{A}=\frac{1}{\sqrt{2}}\left(\xi^{-1} \bm{X}+i \xi \bm{P}\right)
\end{equation}
with now the constant $\xi$ defined as
\begin{equation}
	\xi\equiv\sqrt{\Delta X / \Delta P}
\end{equation}
where $\Delta X$ and $\Delta P$ relate to the resolution of our measuring instrument in position and momentum. We then write the complex parameters $\alpha^i\equiv\alpha^i_3$ and the corresponding state in the Hilbert space simply as
\begin{equation}\label{eq:displacement}
\bm{D}(\alpha) | 0 \rangle =\exp \left\{\alpha^i\, {{\bm{A}}^i}^{\dagger}-{\bar{\alpha}}^i\, {\bm{A}}^i\right\} | 0 \rangle = | \alpha \rangle
\end{equation}
with
\begin{equation}
	\alpha^i=\frac{1}{\sqrt{2}}\left(\xi^{-1} X_0^i+i\, \xi\, P_0^i\right)\ .
\end{equation}
So, this is a Gaussian at position $X_0^i$ and thrown with momentum $P_0^i$. Assuming that the $\bm{X}_A^i$'s with $A=1,2$ are evolving at a much faster pace, we posit that the effective Hamiltonian for the system after measurements of the slow mode is given by~\cite{sahakian2025}
\begin{eqnarray}\label{eq:Halpha}
	\bm{H}_{\alpha} &=& \frac{{\bm{P}^i}^2}{2} + \frac{{\bm{P}_{A}^i}^2}{2} \nonumber\\
	&+&\frac{1}{2} g^2\langle\alpha |
   {\bm{X}^i}^2 | \alpha \rangle {\bm{X}_A^j}^2 -\frac{1}{2} g^2\langle \alpha | \bm{X}_{i}
   \bm{X}_{j}| \alpha \rangle {\bm{X}_A^i} {\bm{X}_A^j} \nonumber\\
   &+& \frac{1}{4} g^2 {\bm{X}_A^i}^2 {\bm{X}_B^j}^2 -\frac{1}{4} g^2 {\bm{X}_A^i} {\bm{X}_A^j} {\bm{X}_B^i} {\bm{X}_B^j} \nonumber \\
   &\equiv & \frac{{\bm{P}^i}^2}{2}  + \bm{H}^{\alpha}_R
\end{eqnarray}
and the state of the system after a position or momentum measurement of the diagonal mode leads to the density matrix given by~(\ref{eq:dm}).
The key here is that the slow mode comes into the fast dynamics adiabatically, so that we take the expectation value of the interaction terms in the coherent state $|\alpha\rangle$.

\section{Numerics}
\label{sub:dynamics}

We want to compute the eigenvalues of $\bm{H}^{\alpha}_R$. First, we diagonalize exactly the quadratic part of~(\ref{eq:Halpha}), leading to the frequencies
\begin{equation}\label{eq:omegas}
	\omega_{i,A} = \left\{\begin{array}{ll}g\,\kappa & i=1 \\ \sqrt{g^2 X_0^2 + g^2 \kappa^2} & i>1\end{array}\right.
\end{equation}
where $A=1,2$ is the color index and $i=1,\ldots, (d-1)$ is the target space index; we have arranged so that the two objects are located as follows:
\begin{equation}
	X^i = 0\ \ \ \mbox{for $i>1$}\ ,\ \ \ X^1=\pm X_0\ ,
\end{equation}
with zero relative momentum.
So, the axes are aligned along the two objects. $\kappa$ is defined as
\begin{equation}
	\kappa = {\sqrt{\frac{(d-1) {\Delta X}}{{2\,\Delta P}}}}\ ,
\end{equation}
related to the precisions with which we measure the momenta and positions of the two objects. The expectation from typical experimental setups is that~\cite{nickwheeler}
\begin{equation}
	\Delta X \Delta P \simeq 1 \Rightarrow \kappa \simeq {\Delta X}{\sqrt{\frac{d-1}{{2}}}}\ .
\end{equation}
In general, we want $\kappa \ll X_0$; we fix $\kappa = X_{0,min}/10$, when $X_{0,min}$ is the smallest separation between the objects we will consider. Typically, we will be exploring the numerics in a region where $\kappa < X_0/100$\footnote{Interestingly, the $\kappa\rightarrow 0$ limit is singular in the Hamiltonian.}.   
Note also that the eigenvalues~(\ref{eq:omegas}) are the same for both color indices $A=1,2$ due to a residual unbroken U(1) gauge symmetry. 

Putting things together, we get the Hamiltonian
\begin{eqnarray}
	\bm{H}^\alpha_R =&& \frac{g^2}{16\,\omega_{i,B}\,\omega_{j,C}}\times \nonumber \\
      &&\Big[
        \bm{\mathcal{A}}_{B,i} \cdot
        \bm{\mathcal{A}}_{C,j} \cdot
        \bm{\mathcal{A}}_{B,i} \cdot
        \bm{\mathcal{A}}_{C,j}
        -\bm{\mathcal{A}}_{B,i} \cdot
         \bm{\mathcal{A}}_{C,j} \cdot
         \bm{\mathcal{A}}_{C,i} \cdot
         \bm{\mathcal{A}}_{B,j}
      \Big]
      \nonumber \\
    &&  + \omega_{i,A}\, \bm{n}_{i,A}\, \delta_{m,n}
        + \frac{1}{2}\,\Omega_i\, \delta_{m,n}\label{eq:HRalpha}
\end{eqnarray}
where\footnote{To increase numerical stability, we subtract the large constant term $(d-1)(d-2)/2\kappa^2$ from the energy computations.}
\begin{equation}
	\bm{\mathcal{A}}_{B,i}\equiv \bm{A}^{\dagger}_{B,i} + \bm{A}_{B,i}\ \ \ \ ,\ \ \ \bm{n}_{i,B} \equiv \bm{A}^{\dagger}_{B,i}\bm{A}_{B,i}
\end{equation}
and
\begin{equation}
	\Omega_i = \left\{\begin{array}{ll}g\,\kappa+\frac{(d-1) (d - 2)}{2\, \kappa^2} & i=1 \\ g X_0 + \frac{(d-1) (d - 2)}{2\, \kappa^2} & i>1\end{array}\right.
\end{equation}
The Gauss constraint takes the form
\begin{equation}
	\left(\bm{A}^{\dagger}_{2,i}\bm{A}_{1,i}-\bm{A}^{\dagger}_{1,i}\bm{A}_{2,i}\right)|\Psi\rangle = 0
\end{equation}
for the residual U(1) gauge symmetry associated with rotating the $1$ and $2$ directions in color space. The other two gauge symmetries are broken explicitly by the arrangement of the objects.

We introduce the dimensionless parameters
\begin{equation}
	g_\kappa \equiv g^{1/3}\,\kappa\ \ \ ,\ \ \ g_X \equiv g^{1/3}\,X_0
\end{equation}
to track the numerical simulation. The task is then to compute as many of the energy eigenvalues for physical states as possible for the operator~(\ref{eq:HRalpha}). Appendix B shows the direct product representation employed in the numerical simulation and some additional technical details. Here, we highlight key elements of the numerical simulation:

\begin{itemize}
	\item {\bf UV matrix cutoff:} For the Hamiltonian diagonalization, we employ the Hilbert space cutoff approach of~\cite{Trzetrzelewski:2003sz,Ambrozinski,Motycka:2014vra}: using the harmonic oscillator basis introduced above, we cut off the spectrum at some fixed level $L$. The Hilbert space is then $L^{2\, (d-1)}$-dimensional, where the factor of two comes from the two colors of off-diagonal excitations\footnote{In general, this is $L^{(N^2-N)(d-1)}$ dimensions for $SU(N)$ -- this is why we fix $N=2$ to manage the computation.}. For $d=3+1$ and $L=10$, we are dealing with matrices that are around a million by a million... The immediate challenge becomes computer memory. To handle this, we employ a trick: we keep all operators in direct product form and encode the Hamiltonian as a linear operator that is never represented as a full matrix. Then, to find eigenvalues and eigenvectors, we employ an iterative algorithm based on the Krylov method to find a fraction of the eigenvalues and eigenvectors. This approach moves the challenge from memory needs to computation speed -- since we then need to do many more matrix multiplications of smaller matrices. To handle this, we employ massive parallelization. In this work, we use a system with 128 CPU cores computing all in parallel. This allows us to produce results for up to $L=9$ for $d=2+1$ dimensions, and $L=6$ for $d=3+1$ dimensions in a reasonable amount of time (one scan of the parameter space takes about 3-4 days). For $d=2+1$, we are able to do exact diagonalization of the Hamiltonian and find all eigenvalues up to the cutoff $L$; for $d=3+1$, we need to use the Krylov technique and get about 10\%-20\% of the spectrum. All this implies that, when we compute the partition function at finite temperature $t$, we cannot push the temperature too high since this renders the results more sensitive to the cutoff. The eigenvalues will scale as $g_X$ at strong coupling, which means our control over the numerics is better at strong coupling than at small coupling. We will refer to this impact of the cutoff $L$ as the {\em UV matrix cutoff} challenge since it affects high temperatures. To track the errors induced by this effect, we increase $L$ by unit steps until $L=6$, then compute a conservative error on the eigenvalues by looking at the fractional change between the last two values of $L$ employed. Since we use the larger $L$, this estimate of the error is larger than the actual accuracy. We also target a 10\% fractional error throughout: if the error on the eigenvalues dips below 10\% at a cutoff $L$, we stop increasing $L$ and move on in the parameter space. We collect in Appendix A figures that quantify all this error analysis.

	\item {\bf Truncation error:} When computing the entropic force, we first compute the free energy (in dimensionless units)
	\begin{equation}
		f = - t\,\ln \mbox{Tr}\, \bm{Z}_\alpha^T
	\end{equation}	
	where $\bm{Z}_\alpha^T$ was defined in~(\ref{eq:Z}). We then construct the entropy from
	\begin{equation}
		S = -\frac{\partial f}{\partial t}\ .
	\end{equation}
	The entropic force becomes
	\begin{equation}
		f_{ent} = t \frac{\partial S}{\partial g_X}\ .
	\end{equation}
	The challenge is that the computation of the partition function is necessarily truncated since we don't have all the eigenvalues
	\begin{equation}
		\mbox{Tr}\, \bm{Z}_\alpha^T = \sum_n e^{-\epsilon_n/t}
	\end{equation}
	where $\epsilon_n$ are the dimensionless eigenvalues: higher $t$ will require inclusion of higher energy levels, which then runs into the cutoff limitation. Furthermore, the eigenvalues will tend to decrease at small $g_X$, making the truncation issue worse. 
	The end result of all this is that we have control over the numerics in a window of the parameter space given by
	\begin{equation}\label{eq:params}
		0.1 < g_X \lessapprox 40\ \ \ ,\ \ \ 3 \lessapprox t \lessapprox 8\ .
	\end{equation}
	given the computer hardware limitations we have. Within this window, errors hover below 10\% for larger $g_X$, and become larger at smaller $g_X$. This fortunately turns out to be enough to capture most of the physics. We collect a more detailed error analysis with quantitative assessments in Appendix A. 
	
	\item {\bf Numerical derivatives}: In analyzing the thermodynamics of the system, we need to compute derivatives using a finite difference method, which introduces additional errors, specially at small $t$.
\end{itemize}

With these limitations in mind, we are able to reliably identify a phase transition or cross-over at $r\sim\ell_P$, emergence of space and gravity beyond that, horizon physics, and even qualitative new insight about the region inside the horizon.

\subsection{$3+1$ dimensions}

\begin{figure}[h]
  \begin{subfigure}{0.48\textwidth}
  \centering
  \includegraphics[width=\linewidth]{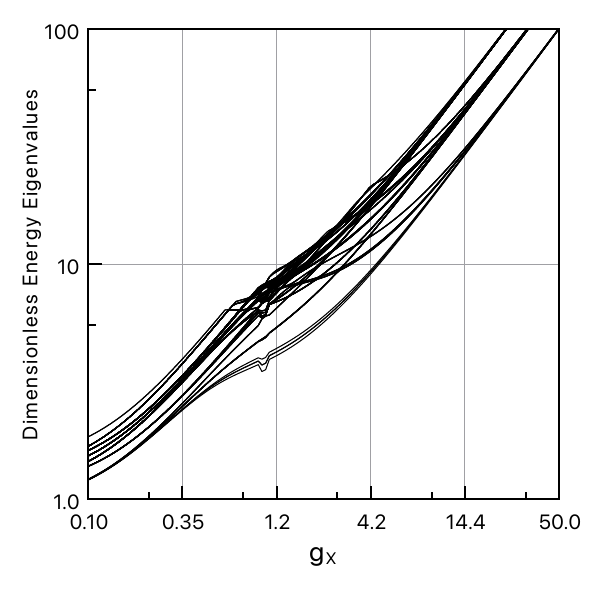}
  \caption{}
  \end{subfigure}
  \hfill
  \begin{subfigure}{0.48\textwidth}
    \centering
  \includegraphics[width=\linewidth]{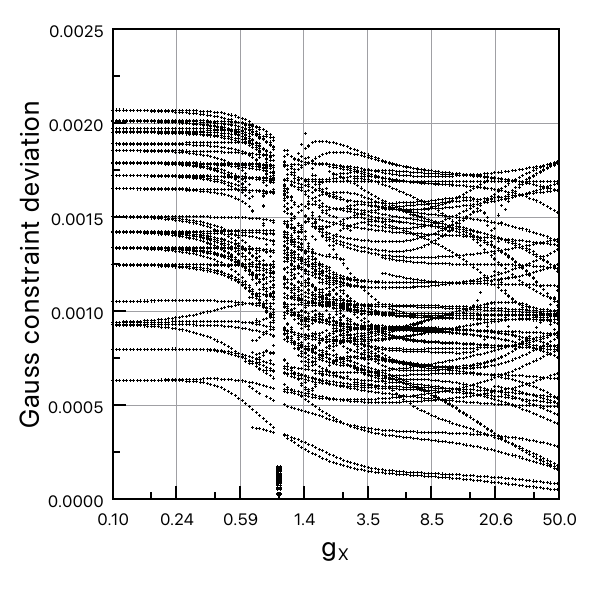}
  \caption{}
	\end{subfigure}
  \caption{\em (a) The low eigenvalues of the Hamiltonian~(\ref{eq:Heff}) on a log-log scale for $d=3+1$. After a regime of eigenvalue crossing, eigenvalue repulsion sets in at higher couplings. (b) Application of the Gauss constraint to eigenvectors. The horizontal scale is logarithmic. In this case, all shown eigenvectors are physical states, lying below the numerical cutoff.}\label{fig:d=3-Eigenvalues}
\end{figure}

Figure~\ref{fig:d=3-Eigenvalues}(a) shows 148 low-lying eigenvalues of the Hamiltonians as a function of $g_X$ -- the coupling in Matrix theory that is also the distance between the objects in M-theory Planck units. We clearly see the weak-strong coupling transition at $g_X\sim 1$, then a regime of eigenvalue crossing, and, when the coupling is large enough, followed by evidence for eigenvalue repulsion characteristic of chaotic systems~\cite{Bohigas}.
To ensure we are considering only physical states, we throw away eigenvectors that are not killed by the Gauss constraint: we do this by computing $\bm{G}|\Psi\rangle$, then finding the standard deviation of the components of the resulting vector $\sqrt{\sum_i v_i^2}/n$; if this is a number greater than a threshold, we throw away the corresponding state. Otherwise, the vector is close enough to the zero vector. The threshold cutoff is chosen consistent with the other numerical errors in the computations, which are independently around 10\%. To be safe, we go a factor of ten lower in identifying the vectors with approximate zero. Figure~\ref{fig:d=3-Eigenvalues}(b) shows the result of applying the Gauss constraint operator on the eigenvectors as a function of coupling. 
Note that, for consistency, if an eigenvector is determined to be unphysical using this criterion {\em at any coupling}, it is thrown away for all couplings. The reliability of the approach is assured by the fact that this criterion is consistent at arbitrarily large coupling, as seen in the Figure. 

\begin{figure}[h]
  \centering
  \includegraphics[width=0.5\columnwidth]{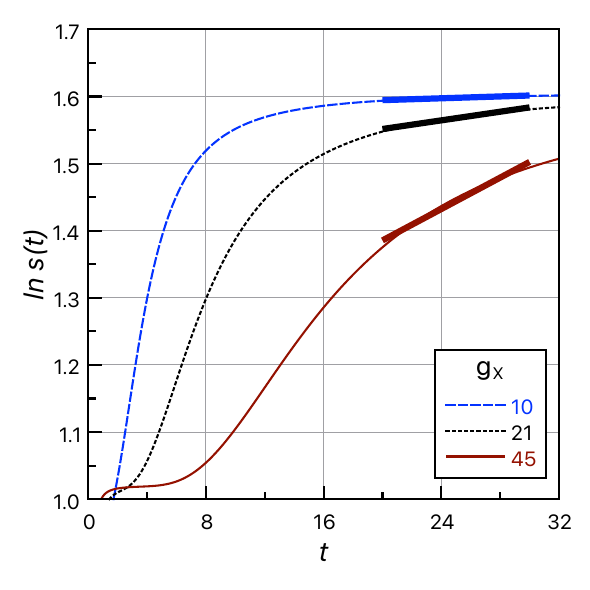}
  \caption{\em The onset of fast scrambling in Matrix theory at large coupling for $d=3+1$.}  \label{fig:energyandscrambling}
\end{figure}
Figure~\ref{fig:energyandscrambling} shows a plot of $\ln S$ versus $t$ at strong coupling. The shape is consistent with fast scrambling scaling given by~\cite{Sekino:2008he,sahakian2025} 
\begin{equation}\label{eq:fastscrambling}
	t\sim g_X\, \ln S_{max}
\end{equation}
at large $g_X$, with $S_{max}$ being the maximum entanglement entropy, {\em i.e.} 
\begin{equation}
	S_{max} = \ln \mbox{min}[D_D, D_R]
\end{equation}
where $D_D$ and $D_R$ are the Hilbert space dimensions of the diagonal and off-diagonal sectors\footnote{In general, for $SU(N)$ Matrix theory, $S_{max}$ scales as $\ln (d-1)\times N$, a number typically much greater than one. For $SU(2)$, this is a number of order one.}. For large $t$, we see this fast scrambling scaling emerging, with additional new interesting features at small $t$.

\subsection{Entropic force}

\begin{figure}[h]
  \centering
  \includegraphics[width=0.9\columnwidth]{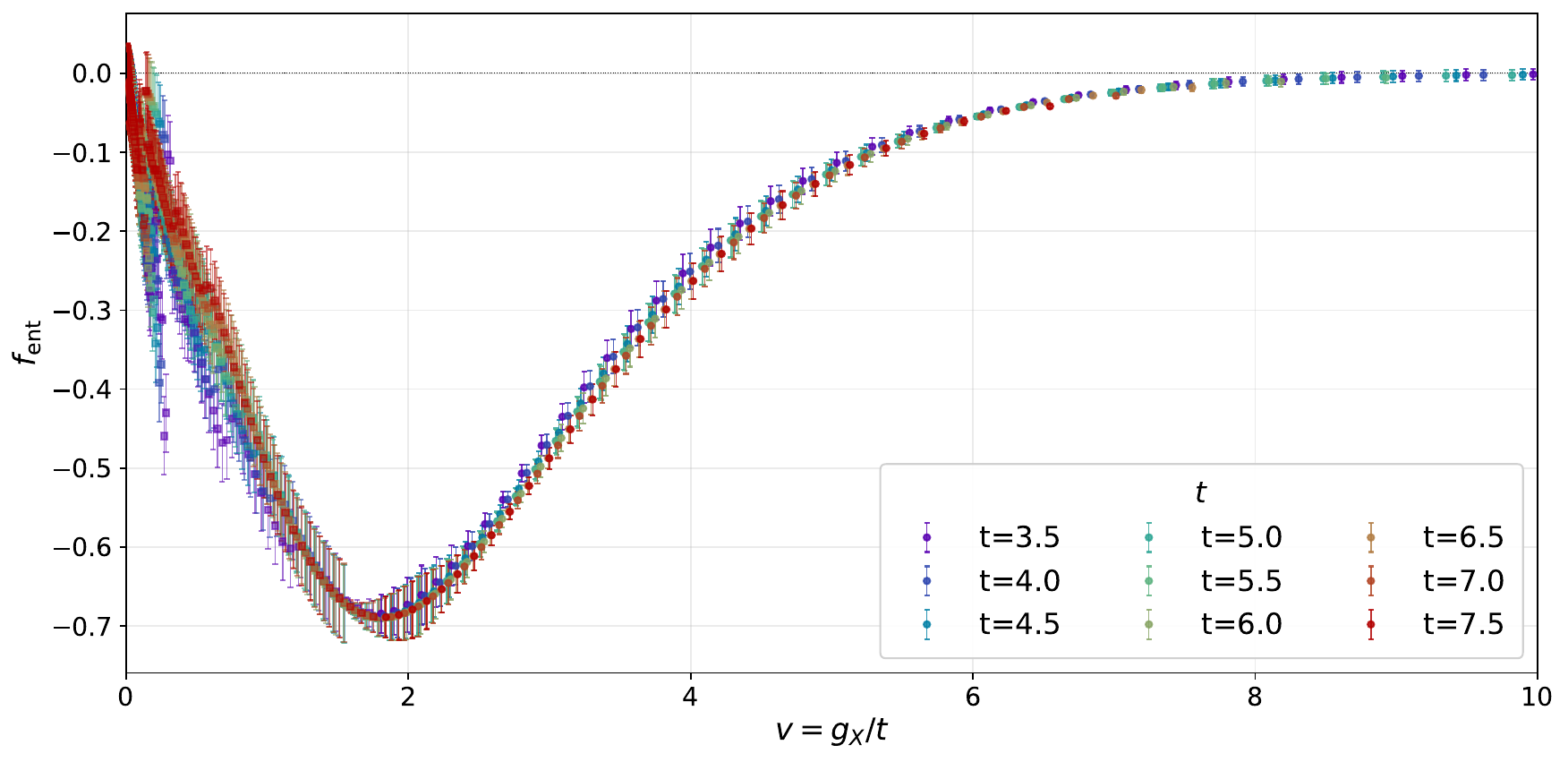}
  \caption{\em The entropic force for $d=3+1$ plotted against $g_X/t$ showing the consistency of the $v=g_X/t$ scaling at large $g_X$, and the breakdown of this parameterization inside the horizon at small $g_X/t$. Error bars shown do not include systematic effects from the matrix UV cutoff and the truncation of the partition function that become important at and below $g_X/t\sim 2$. The systematics in this lower $g_X/t$ region increase the errors to as much as 30\% near the minimum of the force.}
  \label{fig:full-sim-v}
\end{figure}

Figure~\ref{fig:full-sim-v} is a compilation of the main results of this work. We show the computed entropic force for all $g_X$ and $t$ for which we have reliable numerics. The error bars are computed with a nearest neighbor machine learning technique: k-neighbors with $k=30$ for each $g_X/t$. The plot against $g_X/t$ also demonstrates the scaling of the force with respect to $g_X/t$ at large $g_X/t$, but also the breakdown of the dependence of the force on this combination for small $g_X/t$ -- as the points spread around with large error bars. 

\begin{figure}[h]
  \centering
  \includegraphics[width=0.75\columnwidth]{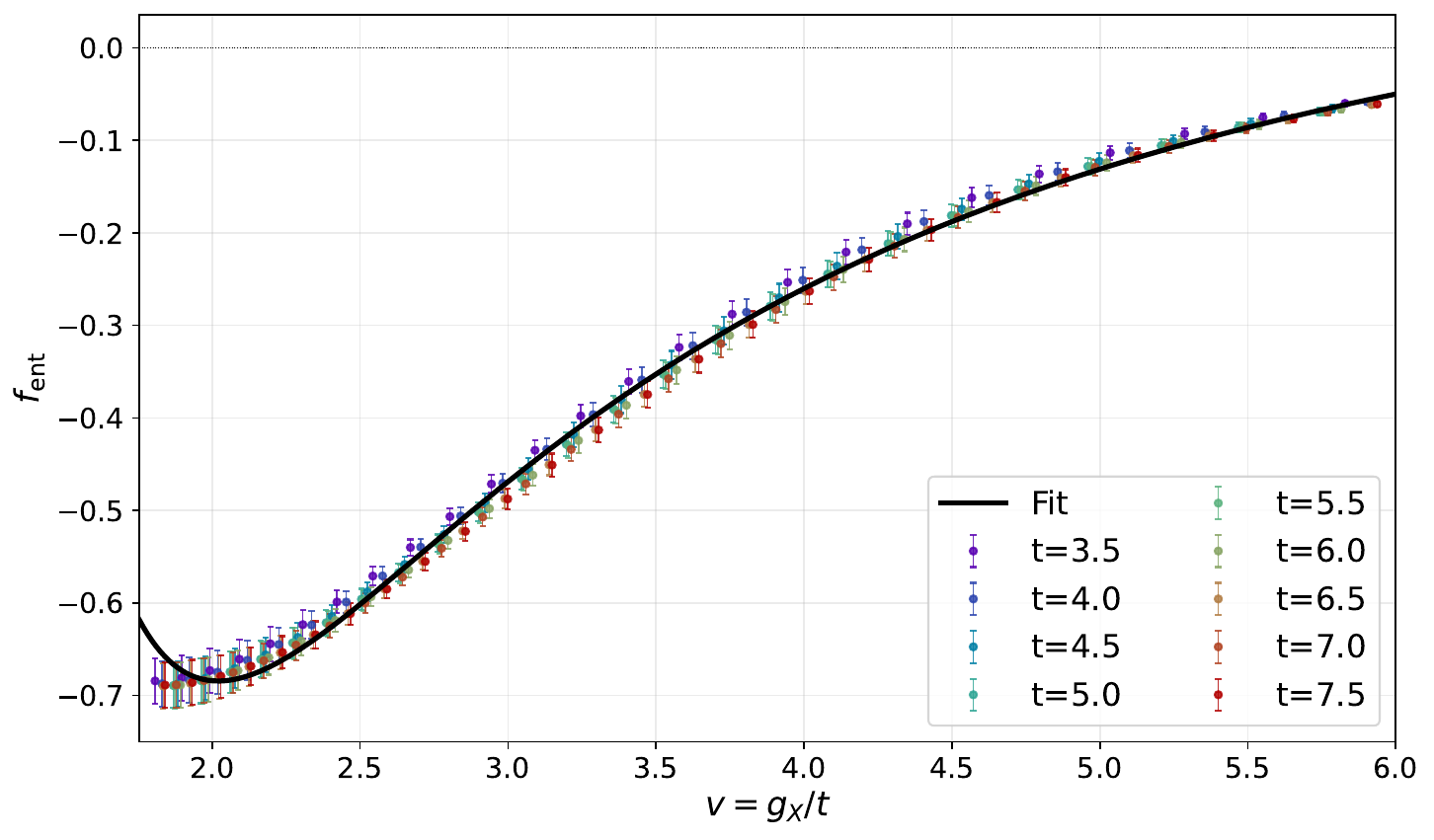}
  \caption{\em Fitting the entropic force outside the horizon for $d=3+1$.}
  \label{fig:force-fit-v}
\end{figure}
Figure~\ref{fig:force-fit-v} zooms onto the region outside the horizon and fits the data using the model given in~(\ref{eq:fit}), with solid statistics indicating the successful identification of gravity -- the leading Newtonian piece plus the general relativistic correction for static objects.

\begin{figure}[h]
  \centering
  \includegraphics[width=0.75\columnwidth]{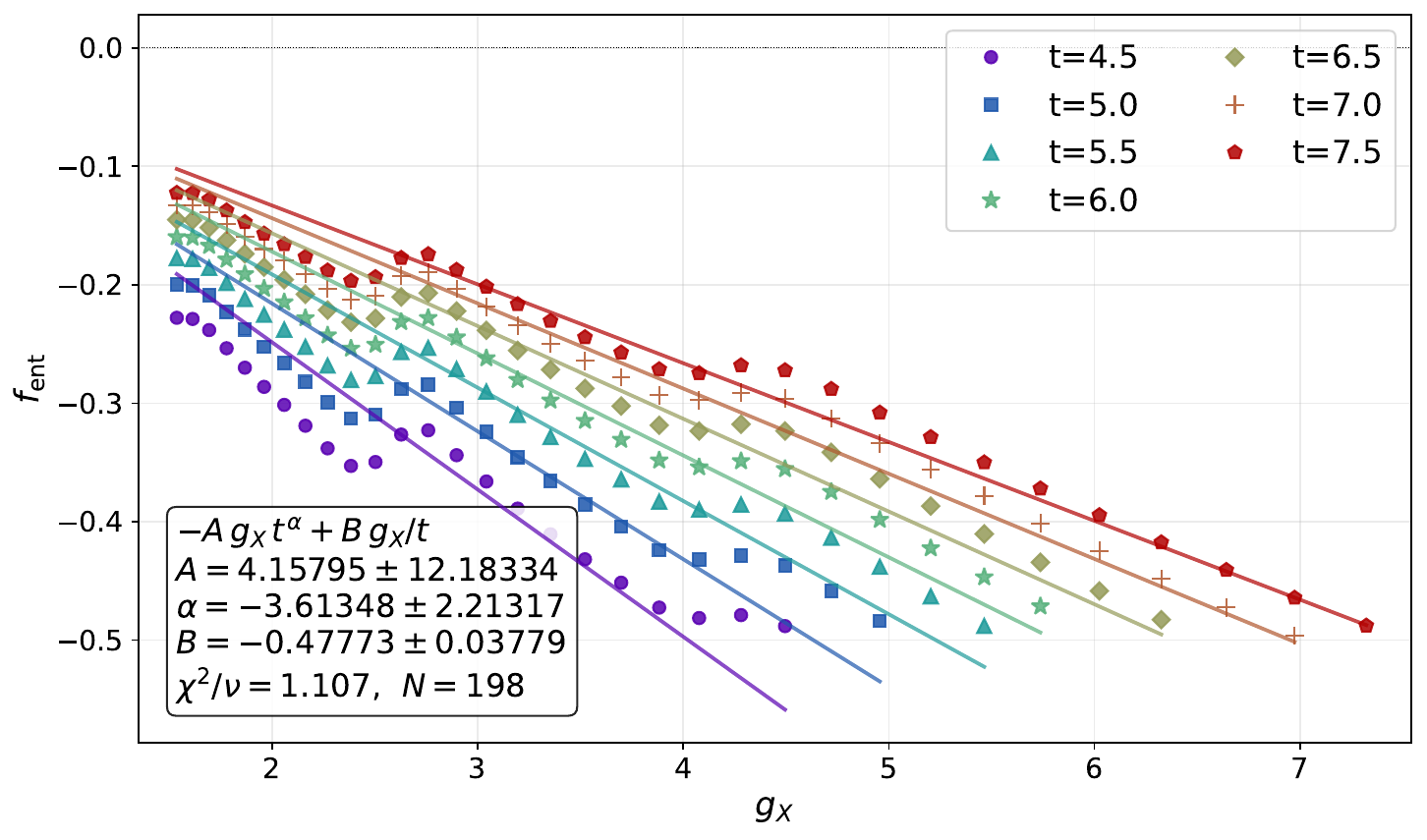}
  \caption{\em Fitting the entropic force inside the horizon for $d=3+1$. Here, we do multiple fits, one for each $t$ given that the $g_X/t$ parameterization fails. Note the layered pattern kinks in the $f\sim -r$ pattern.}
  \label{fig:smallgXfits}
\end{figure}
Next, we focus on the region inside the horizon. Note that the two objects are now intersecting. Figure~\ref{fig:smallgXfits} shows the plot of the entropic force in this region as a function of $g_X$, given that the $g_X/t$ scaling clearly fails as seen from Figure~\ref{fig:full-sim-v}. With an appreciation of the large errors in this region, we employ a fit model 
\begin{equation}
	f_{ent}(g_X,t)=-A\,g_X t^\alpha-B\,\frac{g_X}{t}\ .
\end{equation}
We find
\begin{equation}
	A=4\pm 12\ \ \ ,\ \ \ \alpha=-3.6\pm 2.2\ \ \ ,\ \ \ B=0.48\pm 0.04
\end{equation}
with a reduced $\chi^2_\nu=1.1$ and $N=198$. We can say with reasonable certainty that $f\sim -g_X \sim -r$ given the relatively low error on $B$, a behavior consistent with AdS space inside the horizon.  The errors are clearly large on the other parameters and drawing definitive conclusions beyond this would be speculative. Let us compare this model to the setup where we join the Schwarzschild metric from outside the horizon with AdS space inside -- employing the Barrab\`{e}s-Israel null shell formalism at the horizon~\cite{Barrabes:1991ng,Poisson}. The expected force law inside the horizon becomes
\begin{equation}\label{eq:israel}
	F_{grav}=-16\pi^2 G_4^2 M\,\sigma(M)^2 r+\frac{r}{4\,G_4^2M} 
\end{equation}
where $\sigma(M)$ is the energy density of a wall or membrane at the horizon. With $g_X\sim r$ and $t\sim M$, we can then identify
\begin{equation}
	\sigma(M)\sim M^{(\alpha-1)/2}\ .
\end{equation}
But the error on $\alpha$ is very large. If all the black hole degrees of freedom are to be located at the horizon, we would expect $\sigma\sim 1/M$ or $\alpha = -1$ which is outside the error range we have.

\subsection{$2+1$ dimensions}

In this section, we collect some results for $d=2+1$ dimensions, or two-dimensional target space. Pure gravity is not dynamical in two space dimensions: static massive objects would not feel a force, and we can at best have conical singularities~\cite{carlip2023}. However, the compactified light-cone M-theory we are considering comes with scalars, including from the higher dimensional metric. These typically mediate attractive forces in lower dimensions. We want to see if we can capture some of this physics through the Matrix entropic force in $2+1$ dimensions. 

\begin{figure}[h]
\begin{subfigure}{0.48\textwidth}
  \centering
  \includegraphics[width=\linewidth]{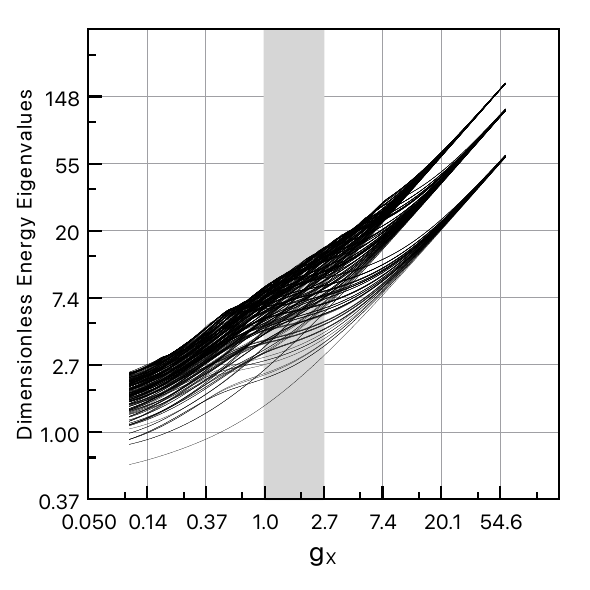}
  \caption{}
\end{subfigure}
\begin{subfigure}{0.48\textwidth}
  \includegraphics[width=\linewidth]{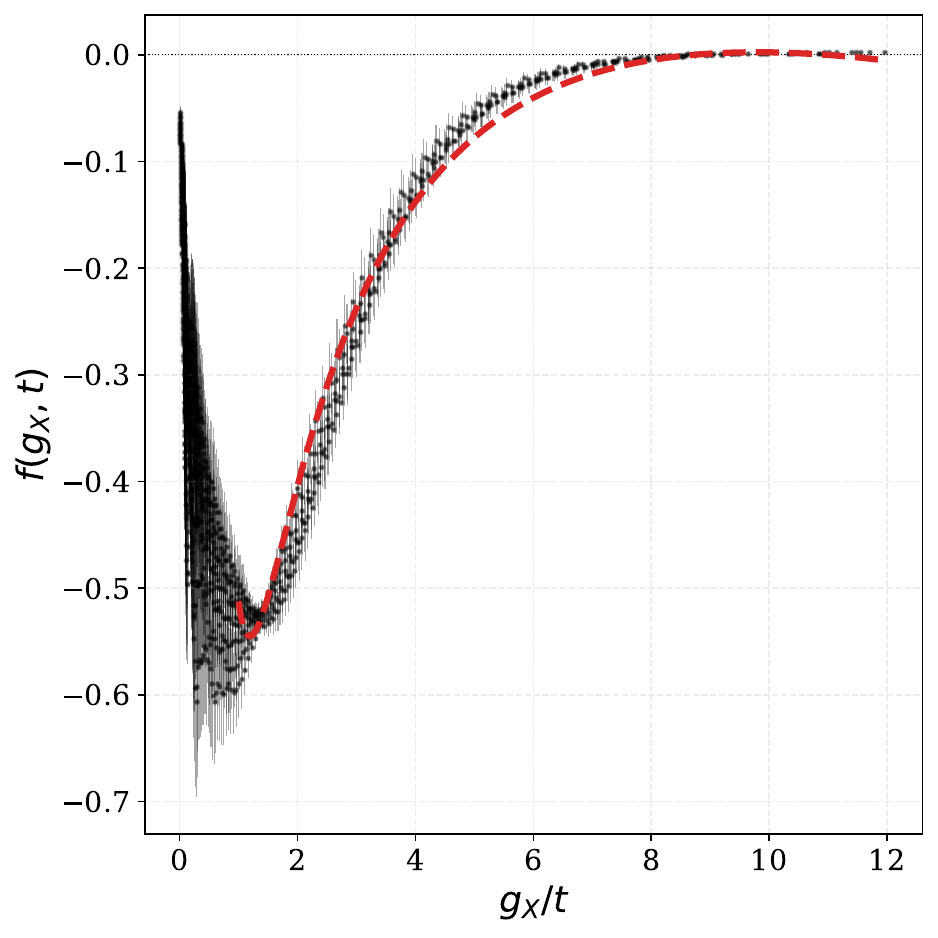}
  \caption{}
\end{subfigure}
\caption{\em (a) Low lying eigenvalues as a function of $g_X$ for $d=2+1$ dimensions. (b) The entropic force for $d=2+1$. }  
\label{fig:d=2-Force}
\end{figure}
In the $d=2+1$ case, the matrices are small enough that we do not need iterative Krylov methods to find a fraction of the eigenvalues and eigenvectors -- we can instead find all of them up to the UV matrix cutoff, which can also be pushed higher. Figure~\ref{fig:d=2-Force}(a) shows the eigenvalues as a function of $g_X$ -- now we are able to have around 500 eigenvalues. We see again a feature at $g_X\sim 1$ and level crossings. 
Figure~\ref{fig:d=2-Force}(b) shows a plot of the entropic force as a function of $g_X/t$ for various $t$. We clearly see an attractive force. The best fit we can find looks like
\begin{equation}
	f_{ent}=c_0+c_1 \frac{t}{g_X} + c_2 \frac{t^2}{g_X^2} + b_1 \frac{g_X}{t}
\end{equation}
with 
\begin{eqnarray}
	c_0&=0.410 \pm 0.004\ \ \ \ \ \ &c_1=-2.19\pm 0.02\nonumber \\
	c_2&=1.28\pm 0.02\ \ \ \ \ \ &b_1=-0.020\pm 0.0002
\end{eqnarray}
with a decent $\chi^2_\nu\sim 2.6$, where we also included a linear term in $g_X$ to capture some of the features inside the `horizon'. Some of the attributes are similar to $d=3+1$ -- an attractive force, a horizon-like feature, more noise inside this horizon. The fall-off with distance now is $1/r$ instead of $1/r^2$. Note however that the scaling of the $1/r$ term in linear with $t$, suggesting possibly a different relation between mass and $t$ at play. We leave the exploration and physical interpretation of these results to another work.

\newpage
\section{Conclusions and Outlook}\label{sec:conclusion}

We have demonstrated that, at strong coupling in {\em bosonic} BFSS  Matrix theory, the gravitational force arises as an entropic force. To maximize entanglement entropy, one can reduce the distance between the two gravitating objects -- smaller $g_X$, or one can increase the masses of the objects -- larger temperature and entanglement parameter $t$. Irrespective, the suggestion is that the notion of space and distance is emergent and entropic. 

In getting at these results, we had to identify the correct operator algebra corresponding to a process of measuring the relative positions of the two objects by an observer at infinity. The assumption was that the chaotic dynamics of Matrix theory in an adiabatic regime results in a density matrix entangling fast off-diagonal modes to slow diagonal ones. An important technical role is played here by the coherent states of finite width associated with the semi-classical measurement process. 

It would help to get better control over numerical accuracy for the region inside the horizon, and hence determine the energy density at the horizon $\sigma(M)$ given by~(\ref{eq:israel}). This will necessitate the inclusion of the fermions in the BFSS theory. In the technique we have adopted, adding the contribution from the fermions is straightforward, void of challenges that are faced by other approaches. The main difference is that the Hilbert space will become larger. On the technical side, this will require code parallelization through the use of modern GPUs given the need to handle larger matrix multiplications. We also need to move from considering hundreds of eigenvalues to thousands, with no more than 10\% error. This can address the numerical ambiguities inside the horizon, but also the role of the constant asymptotic force, and provide control over the systematic errors. As they say, the devil is in the supersymmetry details. We hope to report on this in a future work~\cite{wip}.

\vspace{0.2in}
A few more observations:

\begin{itemize}
	\item An interesting limit is $t\rightarrow 0$ where the entropic force disappears and we have the force
	\begin{equation}
		F \rightarrow -\frac{d\epsilon_0}{dg_X}
	\end{equation}
	where $\epsilon_0$ is the ground state energy of the off-diagonal system. This is indeed how one perturbatively identifies the gravitational force in Matrix theories, provided one includes in the computation of $\epsilon_0$ the contribution from the Matrix theory fermionic modes~\cite{Tafjord:1997bk}-\cite{Lee:2004kv}. Note that the leading contribution from the fermions would exactly cancel the constant term given by $c_0$ in the entropic force~(\ref{eq:fit}) in the limit $t\rightarrow 0$. At finite $t$ -- that is, with gravitating objects that have larger and larger masses, this cancellation is likely less and less perfect, potentially providing a residual long-distance corrections to the gravitational force. This could connect with propositions in the literature for such corrections to explain galactic rotation rates without the use of dark matter for example through Weyl conformal gravity~\cite{Mannheim1989,MANNHEIM_2006,Mannheim_2011}. 
	
	\item Perhaps the most enticing suggestion from this work is the breakdown of general relativity at the horizon, in addition to the replacement of the space behind it with smooth AdS. The space inside the horizon becoming AdS makes the holographic scaling of the black hole entropy with area natural, suggesting a CFT description of the space inside black holes. The entropic force law also has a shape that suggests that degrees of freedom would spend more time near the horizon than in the center, a picture consistent with the fuzzball paradigm~\cite{Lunin_2001,Lunin_2002,Mathur_2005}. We also note that the entropic force demonstrates a layered structure inside the horizon, as seen for example in the close-up of Figure~\ref{fig:smallgXfits}. While this might be a numerical artifact, it might also suggest an onion ring like structure with different negative cosmological constants and spherical layers of AdS. This might also relate to the fact that, at smaller distances, the two objects we are considering will have their horizon intersect. But all this depends on increasing the numerical accuracy of the treatment, to discern physical effects from numerical artifacts. 

	\item We emphasize the remarkable aspect of reproducing the full general relativistic narrative from an entropic force, as suggested first by Verlinde~\cite{Verlinde:2010hp}. This is a highly non-trivial statement, implying that the notion of distance is emergent from entanglement entropy.

	\item In the $2+1$ dimensional case, we have a set of interesting results, an attractive entropic force with similar features to the $3+1$ dimensional case. The only way to reconcile this with a gravitational force is to consider the attractive force from compactified M-theory scalars, which can indeed be geometric in origin. It is however clear that this case requires a separate and special treatment. On the other end, one could expand the computation to $d=4+1$ or higher dimensional M-theory, but this necessitates significantly more computational resources as the Hilbert space dimension grows as $L^{(d-1) N (N-1)}$ for cutoff level $L$, spacetime dimension $d$, and SU(N) Matrix theory.
	
	\item This analysis could be extended to $N>2$, to describe objects with more units of light-cone momenta. This increases the size of the Hilbert space, and large $N$ is prohibitively costly unless one can identify a computational trick that simplifies the setup dramatically. The objects can also be given transverse momenta by expanding around a non-static diagonal matrix configuration. This would be relatively straightforward to implement to capture the gravitational coupling to momentum.

	\item Our setup was purposefully symmetric to simplify the computation: two identical masses momentarily at rest, a fixed distance apart. The mass of the objects can come from wrapping membranes on the compact directions as many times as necessary. But what if we want to capture the gravity between two different masses, say with massses $M_1$ and $M_2$? The first term in equation~(\ref{eq:force}) scales as $M_1\,M_2$ instead of $M^2$, and the second scales as $M_1\,M_2^2$ instead of $M^3$. As a sanity check, we have verified that a $t$ and $t^2$ fit to the data does not work, with a $\chi^2_\nu$ over 100. This suggests that we would need to modify the proposed density matrix~(\ref{eq:dm}) to one that includes two ``temperature'' parameters. A simple extension like $t=\sqrt{t_1 t_2}$ would not work since the second term in the expected force breaks the symmetry between the two masses. In particular, it would be very interesting to figure out the scenario of a large mass and a light probe.
	
\end{itemize}

We end with a final technical note. A great deal of the coding that made this work possible employed the use of Artificial Intelligence (AI) through Anthropic's Claude platform using the Sonnet 4.5 Extended model. While the work can certainly be done without the use of the AI, the speed-up afforded by this tool was significant, the equivalent to 3 to 5 times in human time commitment. This allowed fast iteration in the process of understanding the parameter space and more focus on the physics instead of technical aspects. We believe this approach holds a lot of promise for research in theoretical physics, provided it is used strategically and with proper oversight.

\section{Appendices}\label{sec:appendix}

\subsection{Appendix A: Error analysis}

There are two main sources of error in our numerical treatment, as alluded to in the main text: UV matrix cutoff effect and errors from truncating the partition function. 
\begin{figure}[h]
\begin{subfigure}{0.48\textwidth}
  \centering
  \includegraphics[width=\linewidth]{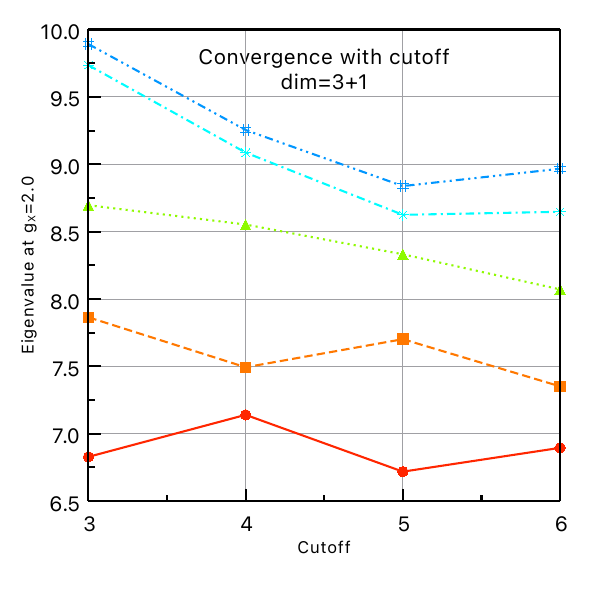}
  \caption{}
\end{subfigure}
\hfill
\begin{subfigure}{0.48\textwidth}
  \centering
  \includegraphics[width=\linewidth]{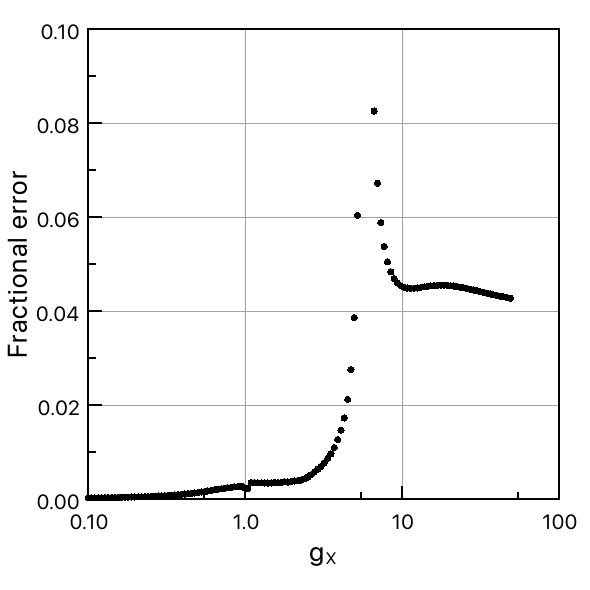}
  \caption{}
\end{subfigure}
   \caption{\em (a) Some low-lying eigenvalues as a function of UV matrix cutoff for $d=3+1$. (b) Fractional error on results arising from the UV matrix cutoff for $d=3+1$ for $t=5$. The horizontal scale is logarithmic. }
 \label{fig:convergence}
 \end{figure}
Figure~\ref{fig:convergence} shows an analysis of errors from the UV matrix cutoff. Figure~\ref{fig:convergence}(a) demonstrates how the eigenvalue computation converges as the cutoff is increased. Note in particular how the higher eigenvalues take more iterations to start converging, limiting the temperature we can raise the system to. Figure~\ref{fig:convergence}(b) shows the fractional error arising from the cutoff effect. This is an upper bound on the error computed from the fractional change between eigenvalues at level $L$ and $L-1$, where $L$ is the largest cutoff used. This error is propagated by quadratures to the partition function and plotted as a function of $g_X$. This is an overestimate of the UV matrix cutoff errors by about a factor of 2 -- given the pattern of convergence we see in~\ref{fig:convergence}(a) and the independent error analysis based on k-nearest neighbors conducted on the whole dataset. We then conclude that the errors are mostly under control (at most around the 6\% level) except interestingly near  $g_X\sim 5$, the horizon point in this $t=5$ case, where we see a spike to around 30\% level. Note also the interesting small dip at the weak-strong coupling transition point $g_X\sim 1$.
  
 \begin{figure}[h]
  \centering
  \includegraphics[width=0.6\columnwidth]{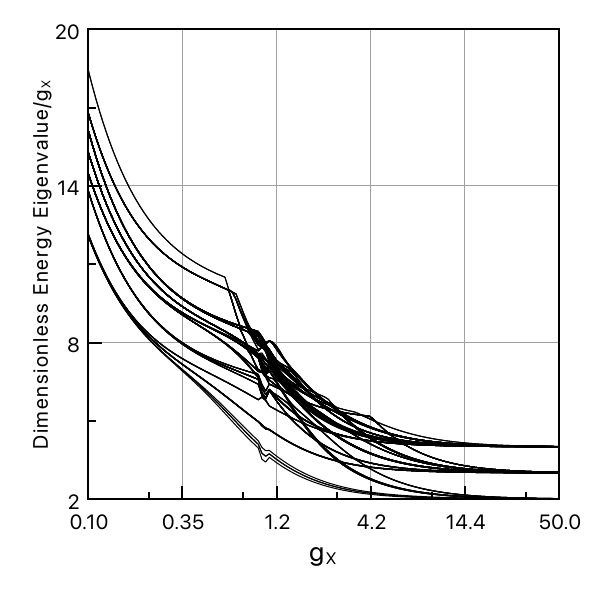}
  \caption{\em Eigenvalues divided by $g_X$ showing that eigenvalues at small $g_X$ are much larger than expected from a geometric series assumption.}
  \label{fig:truncation}
\end{figure}
For $d=3+1$, the truncation of the partition function amounts to around 150 terms out of $6^6=46,656$, a significant effect. We still see truncations errors under control except near the horizon and, more significantly, inside the horizon at smaller $g_X$. This makes our numerics unreliable as we approach $g_X=0.1$, the limit set by the resolution $\Delta X$ of our measurement. However, these errors are a bit too pessimistic for the following reason. The truncated partition function looks like
\begin{equation}
	Z=\sum_{n=1}^k e^{-\epsilon_n/t}
\end{equation}
up to a fixed number $k$. Generally, the fractional error is given by
\begin{equation}
	\mbox{Fractional error}= \frac{\delta Z}{Z\,\ln Z}\ .
\end{equation}
For the error from the UV matrix cutoff, we have by quadrature
\begin{equation}
	\delta_{UV}Z = \sqrt{\sum_{n=1}^k \left(\frac{\delta \epsilon_n}{t}e^{-\epsilon_n/t}\right)^2}\ .
\end{equation}
This is what was done for Figure~\ref{fig:convergence}(b). On the other hand, when we estimate the error in~\ref{fig:truncation}(a), we employ a geometric series approach that assumes harmonic oscillator like structure, $\epsilon_n \sim n\,g_X$, which is an excellent approximation at large $g_X$. This gives the error estimate as
\begin{equation}
	\delta_{trunc}Z = e^{-\epsilon_k/t} \sum_{n=1}^k e^{-n\,(\epsilon_k-\epsilon_{k-1})/t}\ .
\end{equation}
For small $g_X$, Figure~\ref{fig:truncation} shows why this estimate can grossly overestimate the errors: the eigenvalues at small $g_X$ are actually significantly larger than $n\,g_X$, making the partition function truncation more accurate than expected. 

In conclusion, error estimates that provide for internal consistency of the results are dominant outside the horizon region -- using a k-nearest neighbor method, with $k=30$, to compute standard deviation as an estimate for errors on the data points. The main assumption here is that the numerical data is not random, but does capture functional trends and hence must have statistical self-consistency. The fits we presented in the main text adopt this approach for numerical  error estimates. Near the horizon, the matrix UV cutoff dominates what is likely systematic errors; this averages around 30\% at the horizon, and this is what we implemented in the linear fit of $g_{X,min}$ versus $t$. Finally, inside the horizon, the partition function truncation dominates the errors. However, the naive estimate of these errors grossly overestimate them due to a feature in the data that has the ratio of the eigenvalues divided by $g_X$ grow large at smaller $g_X$, making the exponentials in the partition function smaller than expected and, correspondingly, the truncation of the partition function more controllable. Irrespectively, the region inside the horizon still has the largest errors, and we account for this in the analysis of the data.

\subsection{Appendix B: Technical details}

In this appendix, we collect some of the more technical details of the numerical computation. For coding platform, we use Julia on a 128 AMD CPU system with 512Gb memory. The Hamiltonian is represented as a linear operator of a direct products of matrices and the diagonalization is done iteratively using the Krylov technique\footnote{Krylov methods build a low-dimensional subspace by repeated matrix-vector multiplication starting from a probe vector, exploiting the fact that this subspace rapidly captures the extremal spectral components of a matrix. The desired eigenvalues are then extracted by projecting the eigenvalue problem onto this subspace and solving the resulting  smaller system, with the Lanczos algorithms providing orthonormal bases that make the projection numerically stable.}. Each operator is a direct product of $2(d-1)$ $L\times L$ matrices where $L$ is the cutoff. We scan from $L=3$ upward until eigenvalues converge at around the 10\% level. This typically happens around $L=6$ to $L=9$ for the range of parameters $g_X$ and $t$ we need. For $d=2+1$, we can do exact diagonalization of  the Hamiltonian 
\begin{eqnarray}
\hat{H}_3&=& 
   \left( \sqrt{g_X^2+g_\kappa^2}+ g_\kappa\right)\hat{1}{\otimes}\hat{1}{\otimes}\hat{1}{\otimes}\hat{1} \nonumber \\
&+& \frac{1}{8\,g_\kappa\,g_X}\left({\hat{1}{\otimes}\hat{X}^2{\otimes}\hat{X}^2{\otimes}\hat{1}}
   + {\hat{X}^2{\otimes}\hat{1}{\otimes}\hat{1}{\otimes}\hat{X}^2}-2\hat{X}{\otimes}\hat{X}{\otimes}\hat{X}{\otimes}\hat{X}\right) \nonumber \\
   &+& g_\kappa\left(\hat{1}{\otimes}\hat{N}{\otimes}\hat{1}{\otimes}\hat{1}+ \hat{N}{\otimes}\hat{1}{\otimes}\hat{1}{\otimes}\hat{1}\right)+ g_X\left(\hat{1}{\otimes}\hat{1}{\otimes}\hat{1}{\otimes}\hat{N}
   + \hat{1}{\otimes}\hat{1}{\otimes}\hat{N}{\otimes}\hat{1}\right)
\end{eqnarray}
with the Gauss constraint operator
\begin{equation}
\hat{G}_3= -\,\hat{1}{\otimes}\hat{1}{\otimes}\hat{P}{\otimes}\hat{X}
   +\,\hat{1}{\otimes}\hat{1}{\otimes}\hat{X}{\otimes}\hat{P}
 -\,\hat{P}{\otimes}\hat{X}{\otimes}\hat{1}{\otimes}\hat{1}
   +\,\hat{X}{\otimes}\hat{P}{\otimes}\hat{1}{\otimes}\hat{1}\ .
\end{equation}
Here, the $\hat{X}$ and $\hat{P}$ are the $L\times L$ matrix representation of the position and momentum matrices, and $\hat{N}$ is the $L\times L$ number operator.
\begin{equation}
	\hat{X}_{kl} = \sqrt{k}\delta_{l,k+1}+ \sqrt{k}\delta_{l,k-1}\ \ \ ,\ \ \ \hat{P}_{kl} = -\sqrt{k}\delta_{l,k+1}+ \sqrt{k}\delta_{l,k-1}\ \ \ ,\ \ \ \hat{N}_{kl} = k\,\delta_{kl}
\end{equation}

For $d=3+1$, the Hamiltonian becomes
\begin{eqnarray}
\hat{H}_4=&& 
   \left(2\sqrt{g_X^2+g_\kappa^2}+ g_\kappa\right)\hat{1}^{\otimes 6}
   \nonumber\\
&& + \frac{1}{8\,g_\kappa\,g_X}\Big[
      \hat{1}{\otimes}\hat{X}^2{\otimes}\hat{1}{\otimes}\hat{1}{\otimes}\hat{X}^2{\otimes}\hat{1}
   +\,\hat{1}{\otimes}\hat{X}^2{\otimes}\hat{X}^2{\otimes}\hat{1}{\otimes}\hat{1}{\otimes}\hat{1}
   \nonumber\\
&& \hphantom{+\frac{1}{8\,g_\kappa\,g_X}\Big[}
   +\hat{X}^2{\otimes}\hat{1}{\otimes}\hat{1}{\otimes}\hat{1}{\otimes}\hat{1}{\otimes}\hat{X}^2
   +\,\hat{X}^2{\otimes}\hat{1}{\otimes}\hat{1}{\otimes}\hat{X}^2{\otimes}\hat{1}{\otimes}\hat{1}
   \Big]
   \nonumber\\
&& - \frac{1}{4\,g_\kappa\,g_X}\Big[
      \hat{X}{\otimes}\hat{X}{\otimes}\hat{1}{\otimes}\hat{1}{\otimes}\hat{X}{\otimes}\hat{X}
   +\,\hat{X}{\otimes}\hat{X}{\otimes}\hat{X}{\otimes}\hat{X}{\otimes}\hat{1}{\otimes}\hat{1}
   \Big]
   \nonumber\\
&& -\frac{\hat{1}{\otimes}\hat{1}{\otimes}\hat{X}{\otimes}\hat{X}{\otimes}\hat{X}{\otimes}\hat{X}}{4\,g_X^2} \nonumber \\
&& + \frac{1}{8\,g_X^2}\Big[
      \hat{1}{\otimes}\hat{1}{\otimes}\hat{1}{\otimes}\hat{X}^2{\otimes}\hat{X}^2{\otimes}\hat{1}
   +\,\hat{1}{\otimes}\hat{1}{\otimes}\hat{X}^2{\otimes}\hat{1}{\otimes}\hat{1}{\otimes}\hat{X}^2
   \Big]\nonumber\\
&& + g_\kappa\hat{1}{\otimes}\hat{N}{\otimes}\hat{1}{\otimes}\hat{1}{\otimes}\hat{1}{\otimes}\hat{1}
   + g_\kappa \hat{N}{\otimes}\hat{1}{\otimes}\hat{1}{\otimes}\hat{1}{\otimes}\hat{1}{\otimes}\hat{1}
   \nonumber\\
&& + g_X\hat{1}{\otimes}\hat{1}{\otimes}\hat{1}{\otimes}\hat{1}{\otimes}\hat{1}{\otimes}\hat{N}
   \nonumber\\
&& 
   + g_X\hat{1}{\otimes}\hat{1}{\otimes}\hat{1}{\otimes}\hat{1}{\otimes}\hat{N}{\otimes}\hat{1}+ g_X\hat{1}{\otimes}\hat{1}{\otimes}\hat{1}{\otimes}\hat{N}{\otimes}\hat{1}{\otimes}\hat{1}
   \nonumber\\
&& 
   + g_X\hat{1}{\otimes}\hat{1}{\otimes}\hat{N}{\otimes}\hat{1}{\otimes}\hat{1}{\otimes}\hat{1}
\end{eqnarray}
with the Gauss constraint operator given by
\begin{eqnarray}
\hat{G}_4=&& -\,\hat{1}{\otimes}\hat{1}{\otimes}\hat{1}{\otimes}\hat{1}{\otimes}\hat{P}{\otimes}\hat{X}
   +\,\hat{1}{\otimes}\hat{1}{\otimes}\hat{1}{\otimes}\hat{1}{\otimes}\hat{X}{\otimes}\hat{P}
   \nonumber\\
&& -\,\hat{1}{\otimes}\hat{1}{\otimes}\hat{P}{\otimes}\hat{X}{\otimes}\hat{1}{\otimes}\hat{1}
   +\,\hat{1}{\otimes}\hat{1}{\otimes}\hat{X}{\otimes}\hat{P}{\otimes}\hat{1}{\otimes}\hat{1}
   \nonumber\\
&& -\,\hat{P}{\otimes}\hat{X}{\otimes}\hat{1}{\otimes}\hat{1}{\otimes}\hat{1}{\otimes}\hat{1}
   +\,\hat{X}{\otimes}\hat{P}{\otimes}\hat{1}{\otimes}\hat{1}{\otimes}\hat{1}{\otimes}\hat{1}
\end{eqnarray}
The computation is made feasible because of parallelization of the matrix multiplications, running 128 parallel processes simultaneously. Claude AI was used to optimize the code for maximum performance.

The code used to generate all this data can be made available to researchers upon request.

\section*{Acknowledgments}

This work was supported by NSF grant number PHY-2109420 and the Burton Bettinger fund. Anthropic's Sonnet 4.6 Extended was used in generating code and verifying the data analysis.

\section*{References}



\end{document}